\documentclass[twocolumn,showpacs,prl,aps,10pt]{revtex4-1}

\usepackage{graphicx} 
\usepackage{subfigure}
\usepackage{color}
\usepackage{amsmath}
\usepackage{amssymb}
\usepackage{wasysym}
\usepackage{dcolumn}

\begin{document}

\title{The mechanical response of a creased sheet}

\author{F. Lechenault$^{1}$}
\email{frederic.lechenault@lps.ens.fr}
\author{B. Thiria$^{2}$}
\author{M. Adda-Bedia$^{1}$}

\affiliation{$^{1}$Laboratoire de Physique Statistique, Ecole Normale Sup\'erieure, UPMC Paris 6, Universit\'e Paris Diderot, CNRS,
24 rue Lhomond, 75005 Paris, France}
\affiliation{$^{2}$Physique et M\'ecanique des Milieux Het\'erog\`enes, ESPCI ParisTech, UPMC Paris 6, Universit\'e Paris Diderot, CNRS, 10 rue Vauquelin, 75005 Paris, France}

\date{\today}

\begin{abstract}

We investigate the mechanics of thin sheets decorated by non-interacting creases.  The system considered here  consists in parallel folds connected by elastic panels. We show that the mechanical response of the creased structure is twofold, depending both on the bending deformation of the panels and the hinge-like intrinsic response of the crease. We show that a characteristic length scale, defined by the ratio of bending to hinge energies, governs whether the structure's response consists in angle opening or panel bending when a small load is applied. The existence of this length scale is a building block for future works on origami mechanics.

\end{abstract}


\maketitle

Low-dimensional elastic structures such as thin plates and rods routinely undergo large strains whereby a complex network of plastic zones emerges as an outcome of the conjunction between extreme deformations and physical self-avoidance~\cite{witten2007}. Such situations are very common in the living world where, for example, the shape of insect wings~\cite{haas1996} and petal leaves~\cite{couturier2009} is, to some extent, influenced by the protective container which constrains their growth and results in the formation of permanent folds. The physics of folding also arises at small scales, such that in proteins formation or biological membranes. In man-made systems, the question of fold creation is of paramount importance as it arises in a wide spectrum of, apparently disconnected, human achievements ranging from art to space exploration. For example, fashion designers would gain from a better and more systematic understanding of how a particular cut and fold geometry of their fabric may affect the overall shape of a dress~\cite{holdaway1960behavior}. Similarly, mechanical engineers looking to design solar panels or foldable lunar bases need precise origami patterns which not only optimize the logistics and storage when the structural components are folded, but also ensure their safe and reliable tear-free deployment~\cite{miura1985,miura1993,gruber2007}.

In defiance of these enchanting associations, the prototypical system which allows a more controlled study of folds creation and of their mechanical properties assumes a far more undistinguished form: namely a crumpled paper~\cite{witten2007,deboeuf2013}. Careful observations have revealed that decoration of  a thin plate by a crease network dramatically modifies the global mechanical response of the plate resulting, for instance, in aging of the structure~\cite{thiria2011relaxation}, negative Poisson ratio and unusual response to bending and stretching~\cite{Wei2013}. Clearly, the opening and closing of folded objects critically depends on the mechanical characteristics and on the geometrical network of the creases~\cite{ Dias2012,Kang2014,Dias2014}. At the first level of description, the pattern of folds can be modeled by elastic hinges of specific stiffness where each crease lies at the intersection between two rigid, non-deformable panels. Hopefully, the whole mechanism could then be understood by studying the kinematics of geometrically interdependent rigid panels coupled to each other via the spatial distribution of the fold skeleton~\cite{Wei2013,haas1996}. However, this general ``geometrical" description dismisses the role played by the intrinsic elastic response of the panels during the folding and deployment of the structure. Moreover, the hinge rigidity clearly should not be assumed independently (\textit{a posteriori}) but rather be directly deduced from the material properties from which the crease is created.

In this Letter, we provide a complete description of the mechanical response of a folded elastic object under external load.  For the sake of clarity, our experimental systems are one-dimensional and the network of creases consists of parallel folds separated by a typical inter-panel length-scale.  This set-up aspires to realistically imitate the simple 1 fold-2 panels sub-system.  The Letter culminates with the proposition that there exists a critical length governing the mechanical response of the crease that depends both on the panel deformation and the hinge properties. We emphasize that this emerging behavior cannot be captured if the elasticity of the panels is not considered.

\begin{figure}[ht]
\begin{center}
\includegraphics[width=0.48\textwidth]{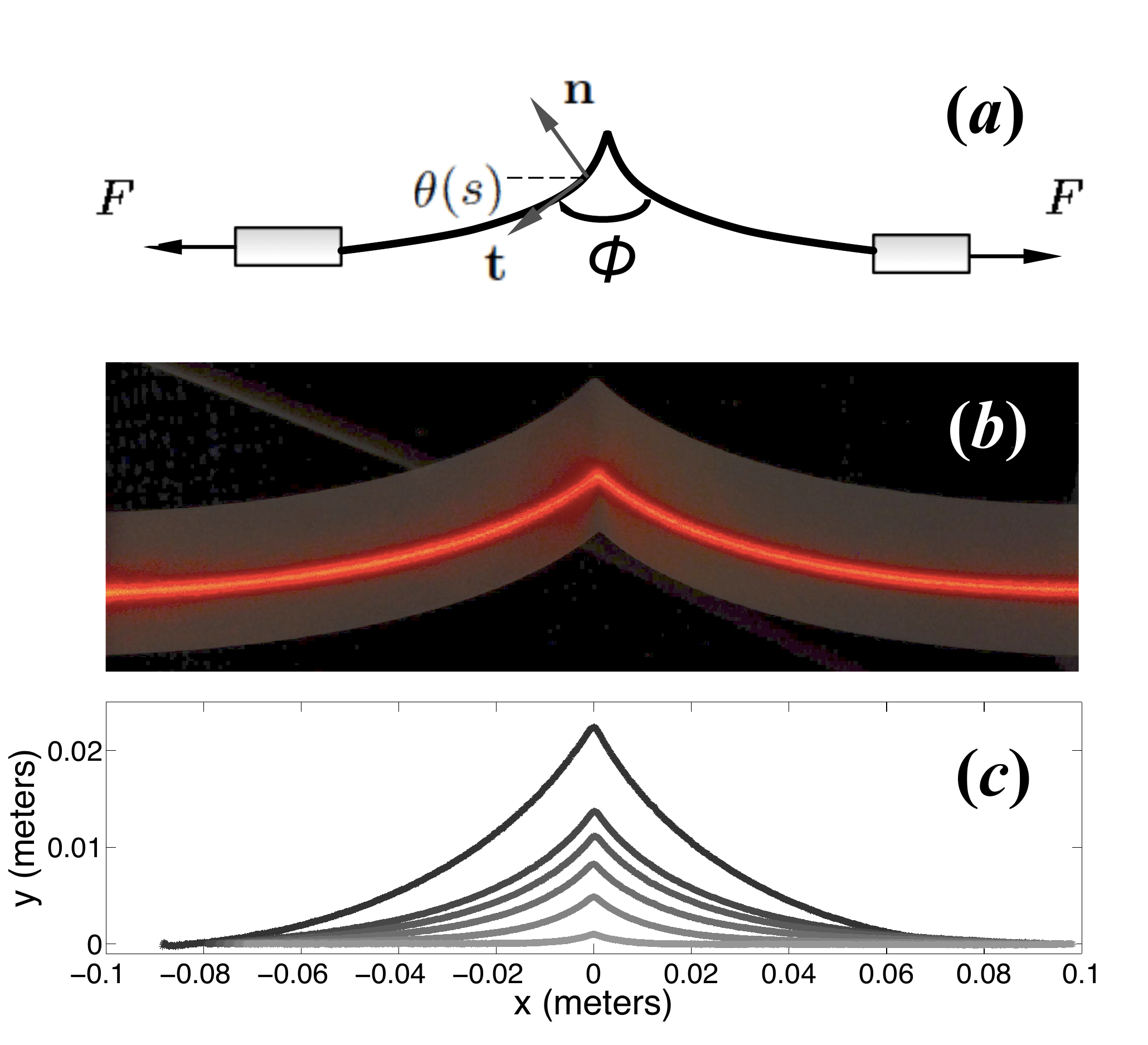}
\caption{a) Schematic of the experimental setup. b) Picture of the laser-illuminated sheet showing the resulting deformation of the fold. c) Profiles $y(x)$ extracted from the pictures for various elongations applied to a sheet of thickness $h=350 \mu m$.}
\label{fig:profile}
\end{center}
\end{figure}

We consider long rectangular Mylar sheets shaped into a single fold. The sheets are $200$mm long, $30$mm wide, and we have investigated three values for their thickness, namely $150\mu m$, $350\mu m$ and $500\mu m$. The crease is produced by loading a hand-shaped pre-crease with a $10$kg weight for $20$ minutes, and then letting it relax for an hour, which is sufficient to
obtain a quasi-stationary state for the fold \cite{thiria2011relaxation}. We will further justify this protocol in the sequel.
In their rest state, the sheets exhibit two flat panels forming an angle $\phi_0$. We then fasten one end of one strip to a fixed clamp and the other end to the probe of an Andilog$^\circledR$ dynamometer in such a way that both ends lie in the same vertical reference plane (Fig.~\ref{fig:profile}(a)). The sample is illuminated by a laser sheet parallel to the long dimension of the strip and perpendicular to the reference plane (Fig.~\ref{fig:profile}(b)). The dynamometer is mounted on a translation stage, allowing to pull on the strip along its length. For each position of the stage, a picture of the illuminated strip is taken with a Nikon$^\circledR$ digital camera, at a $60^{\circ}$ angle with respect to the direction of the light source. The strip's profiles $y(x)$ are extracted from these images (Fig.~\ref{fig:profile}(c)) and correlated to the applied force of the dynamometer. 

In order to extract the full mechanical behavior of the folded thin sheet, we study a limiting case for which {Euler's elastica} yields analytical solutions for the deformation of the panels: a strip creased in the middle of its long dimension, parallel to the short dimension, and pulled from both ends parallel to the strip's flat state as shown in Fig.~\ref{fig:profile}(a).  
The two-dimensional elastica problem to solve is the following~\cite{thiria2011relaxation}
\begin{equation}
BW\theta_{ss}-F \sin\theta=0
\label{eq:elastica}
\end{equation}
where $s$ is the curvilinear coordinate along the profile starting from the crease, $\theta$ its local angle with respect to the horizontal axis, subscript $s$ is the curvilinear derivative, $B=Eh^3/12(1-\nu^2)$ is the bending rigidity of the sheet ($E$ is the Young's modulus and $\nu$ the Poisson ratio), $h$ its thickness and $W$ its width. The boundary conditions read
\begin{equation}
\theta(\infty)=0\;, \quad \theta_s(\infty)=0\;, \quad \theta (0)=\frac{\pi-\phi}{2}
\end{equation}
where $\phi$ is the current opening angle of the fold. Here the term $\infty$ signifies that the length of the elastica is large compared to any other length scale in the problem. Solving this equation leads to
\begin{equation}
\tan\frac{\theta}{4}=\tan\frac{\theta(0)}{4}\,\exp{-\frac{s}{c}}
\label{eq:exp}
\end{equation}
where $c$ is the elastic length scale of the problem~: 
\begin{equation}
c=\sqrt{\frac{BW}{F}}
\label{eq:c}
\end{equation}
Now, we use this exact result  to extract the opening angle of the fold $\theta(0) $ and the characteristic length scale $c$ in a realistic experimental situation, in order to obtain the response of the crease to the loading moment. For this, one needs to transform the profiles $y(x)$ of Fig.~\ref{fig:profile}(c) into a parametrization $\theta(s)$.

\begin{figure}[ht]
\begin{center}
\includegraphics[width = 0.48\textwidth]{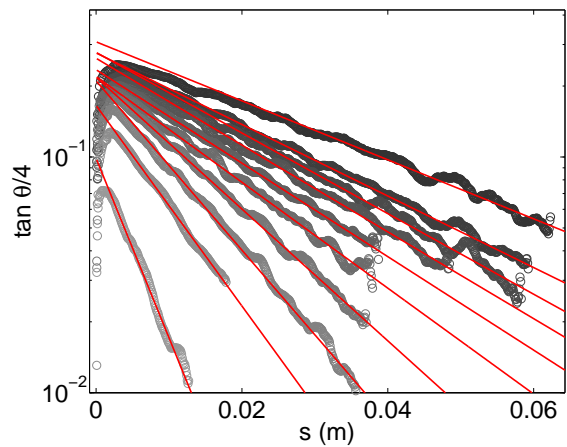}
\caption{Lin-log plot of $\tan\theta/4$ as function of the curvilinear coordinate $s$ for a representative set of elongations corresponding to the profiles shown in Fig.~\ref{fig:profile}(c), and their respective exponential fits (red lines). The darker the dots, the smaller is the force. The oscillations at the tails of each curve are due to the unavoidable errors in the computation of the slopes $y'(x)$.}
\label{fig:lintan}
\end{center}
\end{figure}

\begin{figure}[ht]
\begin{center}
\includegraphics[width = 0.23\textwidth]{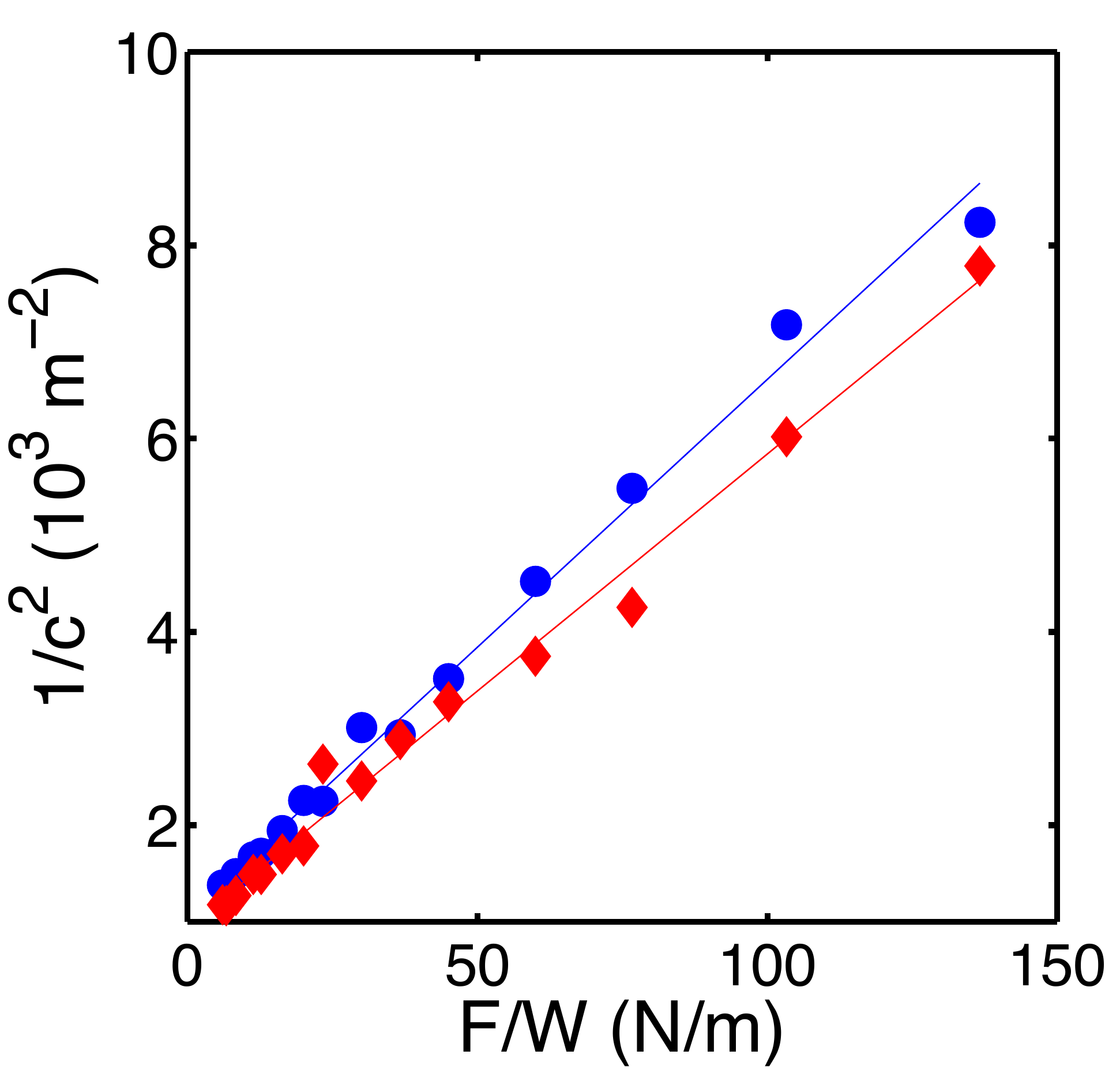}
\includegraphics[width = 0.23\textwidth]{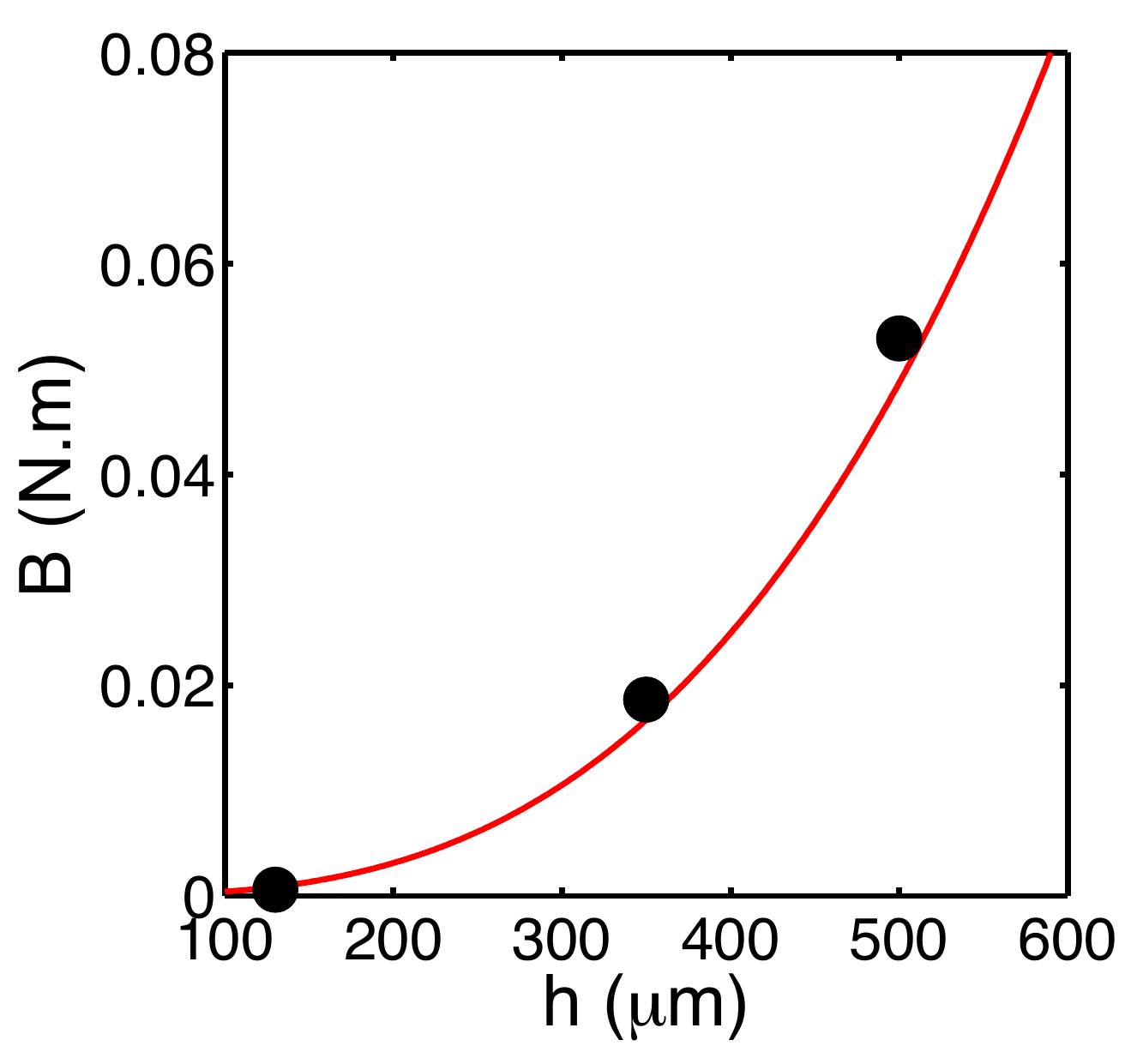}\\
\includegraphics[width = 0.23\textwidth]{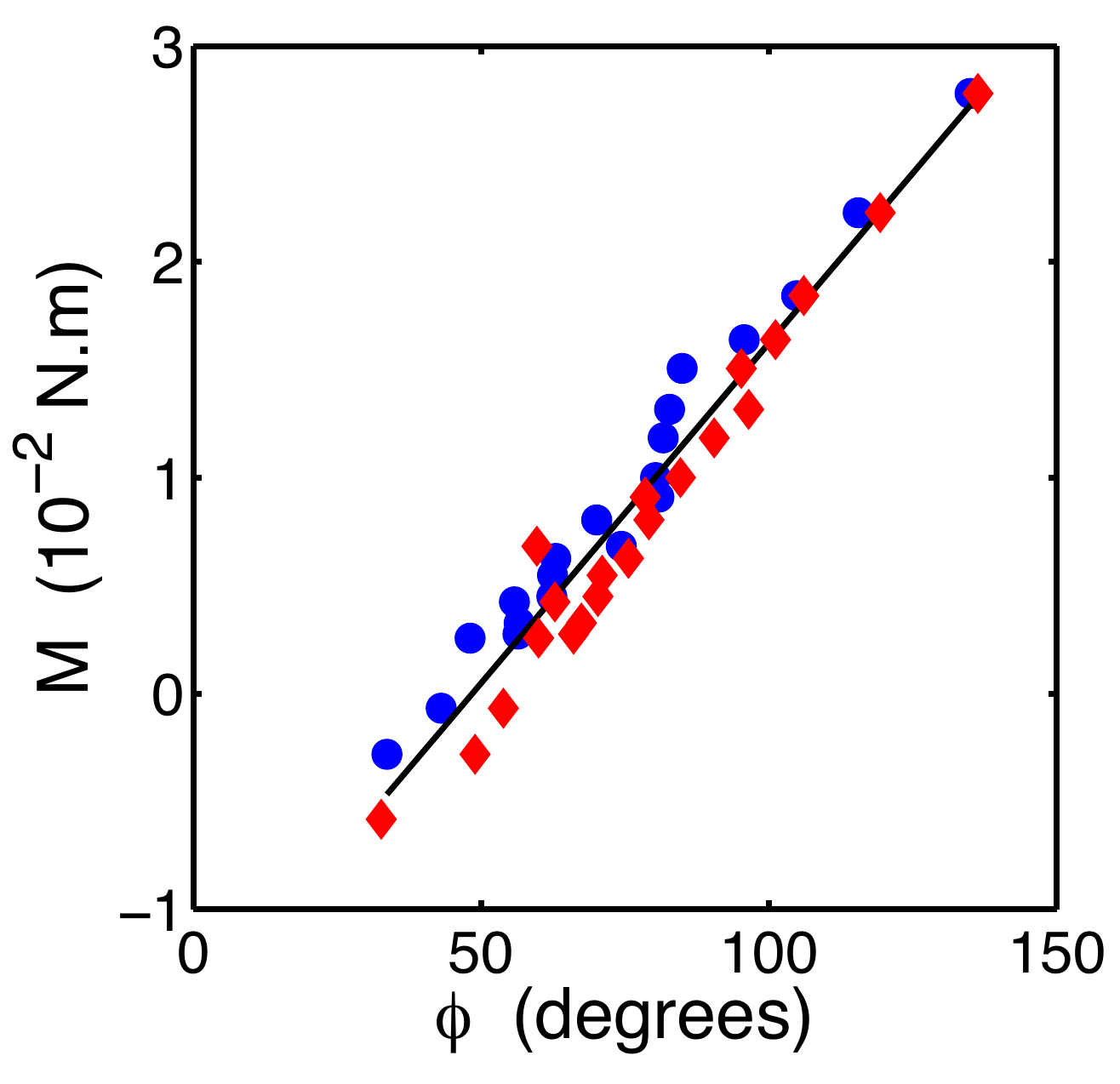}
\includegraphics[width = 0.23\textwidth]{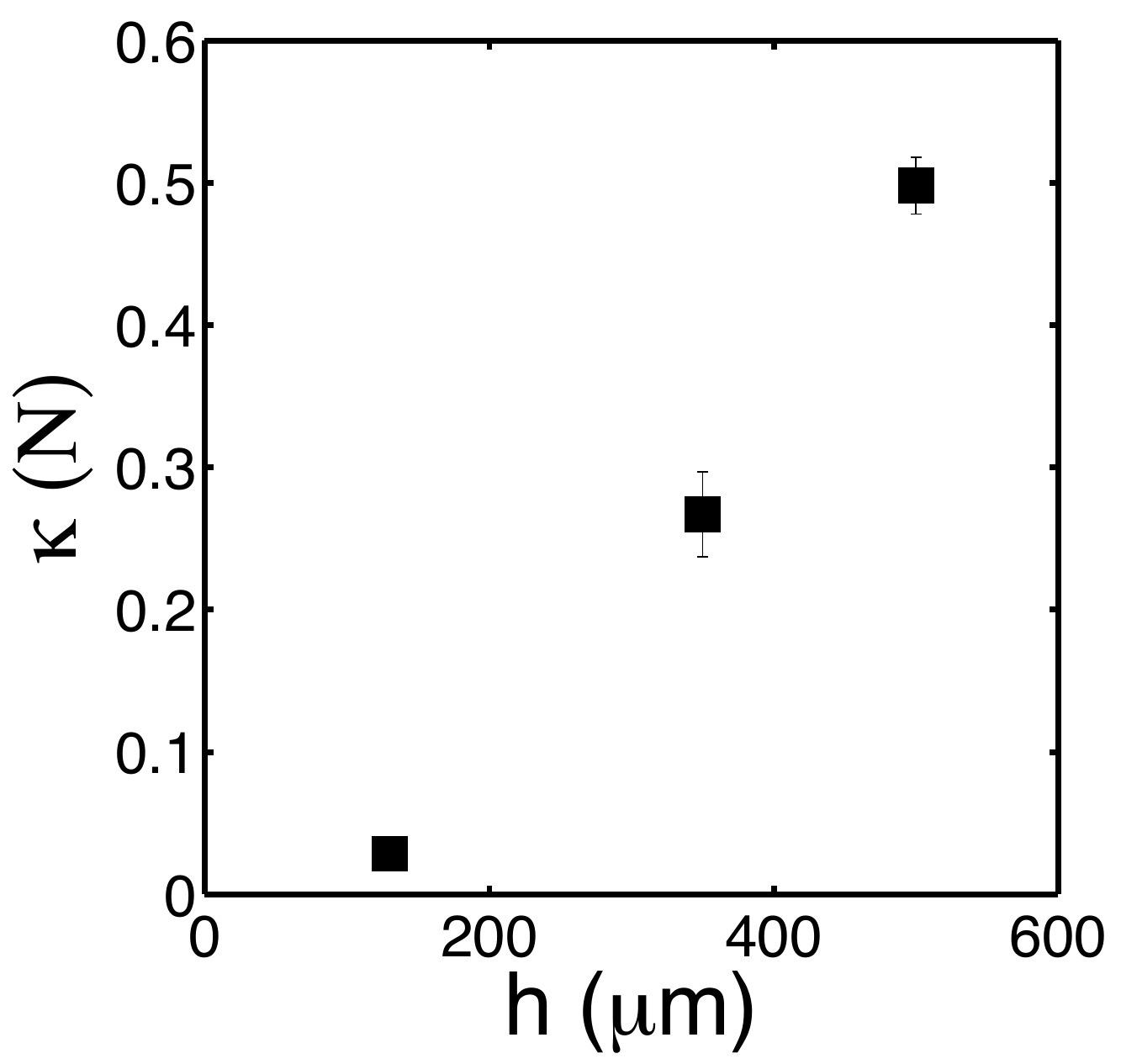}
\caption{a) $1/c^2$ function of the applied force (normalized by the width of the strip) for right portions (blue disks) and left portions (red diamonds) of the strip, for a sheet of thickness $h=350 \mu m$. Solid lines are corresponding linear fits. b) The bending modulus dependence on material thickness $h$. The dashed line is the function $B=E h^3/12(1-\nu ^2)$, with $E=4. 10^9 Pa$ and $\nu=0.38$ extracted from the tabulated value for Mylar sheets. c) Constitutive mechanical behavior of the crease: applied moment $M$ function of the opening angle $\phi$ for right portions (blue disks) and left portions (red diamonds) of the strip of the same sheet as in (a). Solid line is a global linear fit. d) The rigidity $\kappa$ of the crease for different thicknesses. The error bars are $95\%$ confidence intervals.}
\label{fig:Fig3}
\end{center}
\end{figure}

Fig.~\ref{fig:lintan} shows that the shape of the side panels follows the exponential decay predicted by Eq.~(\ref{eq:exp}) for a large range of $s$ and breaks down in the vicinity of the fold. This latter behavior is anticipated because the fine structure of the fold should show off at scales comparable to the sheet thickness. The characteristic length scale $c$ can be extracted from these exponential decays. Fig.~\ref{fig:Fig3}(a) shows that $1/c^2$ is a linear function of the applied force, as predicted by Eq.~(\ref{eq:c}). Fig.~\ref{fig:Fig3}(b) shows the bending modulii of three samples with different thicknesses extracted from these measurements and compared with the definition $B=E h^3/12(1-\nu ^2)$ of the bending modulus. The observed quantitative agreement validates the assumptions put forward in the derivation of Eq.~(\ref{eq:exp}).

A valuable byproduct of the previous fit is the current opening angle of the crease $\phi = 2(\frac{\pi}{2}-\theta (0))$ as function of the applied force. If the crease can be considered as a torsional hinge, one expects a well defined constitutive behavior relating the applied moment to the angle difference of the loaded and rest states of the crease. Fig.~\ref{fig:Fig3}(c) shows the opening moment $M$, evaluated by multiplying the pulling force by the height of the profile, as a function of the crease angle $\phi$. After one loading/unloading cycle, this response becomes reversible. In this reversible regime, the moment is linear in the opening angle, thus allowing to characterize the crease rigidity by a single parameter $\kappa$ defined by $M = \kappa W (\phi-\phi_0)$ where $\phi_0$ is the rest angle of the crease. The linear fits for the three studied samples $h=130 \mu m, 350 \mu m$ and $500 \mu m$ yield respectively $(\kappa,\phi_0)=(0.029\pm0.007 N,27^{\circ})$, $(0.27\pm0.03 N,43^{\circ})$ and $(0.5\pm0.02 N,39^{\circ})$ (see Fig.~\ref{fig:Fig3}(d)).

\begin{figure}[ht]
\begin{center}
\includegraphics[width = 0.48\textwidth]{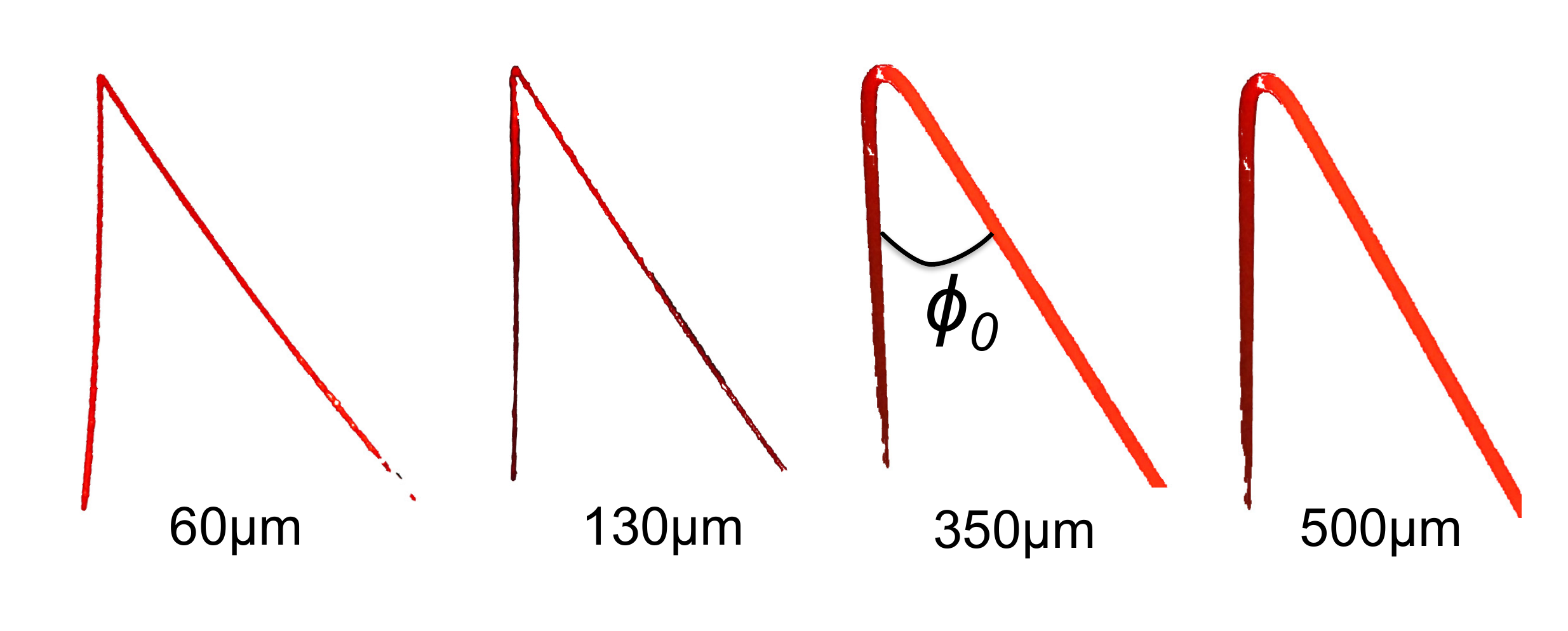}
\caption{Spontaneous relaxation of folds in Mylar sheets of various thicknesses produced by the same protocol. The images show that, for the same material, the final rest angles marginally depend on the sample thickness.}
\label{fig:relf}
\end{center}
\end{figure}

Let us comment on this constitutive behavior starting with the rest angle $\phi_0$. The protocol for crease creation, by heavily loading a pre-shaped crease with plates parallel to its faces, results in a typical rest angle of the fold that depends weakly on the sheet thickness. This simple protocol leads to reproducible results and, in turn, governs the mechanical properties of the hinge. This feature can be understood qualitatively by noticing that the creation of the crease involves localized plastic deformations, through localized storage of bending elastic energy~\cite{deboeuf2013}. Assuming an ideally plastic behavior of the material, and considering the thickness $h$ as the crease characteristic radius of curvature, the plastic strain scales as $\epsilon_p\sim h\phi_{p}/h=\phi_p$, where $\phi_p$ is the characteristic angular region in which the irreversible deformations are localized. Using the constitutive relation  that relates the plastic strain to the yield stress $ \sigma_Y$, one deduces that $\phi_p\sim\epsilon_p \sim \sigma_Y/E$ independently of the sample thickness. If one assumes that the opening angle $\phi_0$ is a univocal function of $\phi_p$, this simple estimate shows that $\phi_{0}$  depends on the mechanical properties of the material but not on its thickness. In order to test this result, we prepared additional samples with various thicknesses by using the same protocole, and found that the rest angles always lie between $30^{\circ}$ and $40^{\circ}$ with no significant dependance on $h$ (see Fig.~\ref{fig:relf}). Notice that these rest angles are smaller than those obtained from the mechanical tests carried out above. This might be explained by noticing that the crease is subjected to long relaxation processes~\cite{thiria2011relaxation} and the mechanical testing might accelerate the convergence to the final rest angle. This is supported  by the quick convergence of our loading cycles to a well defined reversible response.

\begin{figure}[ht]
\begin{center}
\includegraphics[width = 0.4\textwidth]{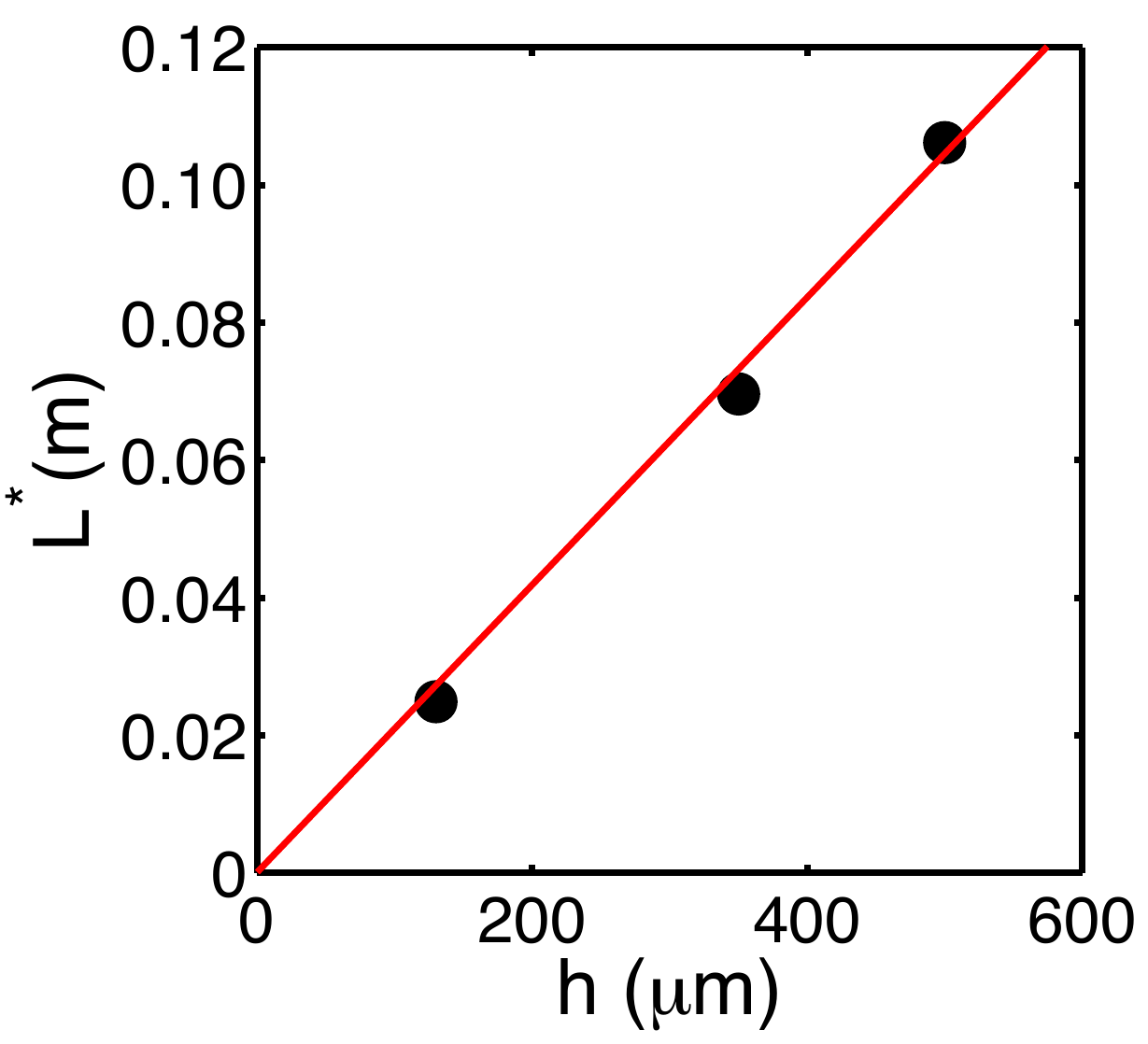}
\caption{Origami length scale $L^*$ as defined in the text, for different sheet thicknesses. The solid line is a linear fit.}
\label{fig:Fig5}
\end{center}
\end{figure}

\begin{figure}[ht]
\centering
\includegraphics[width = 0.48\textwidth]{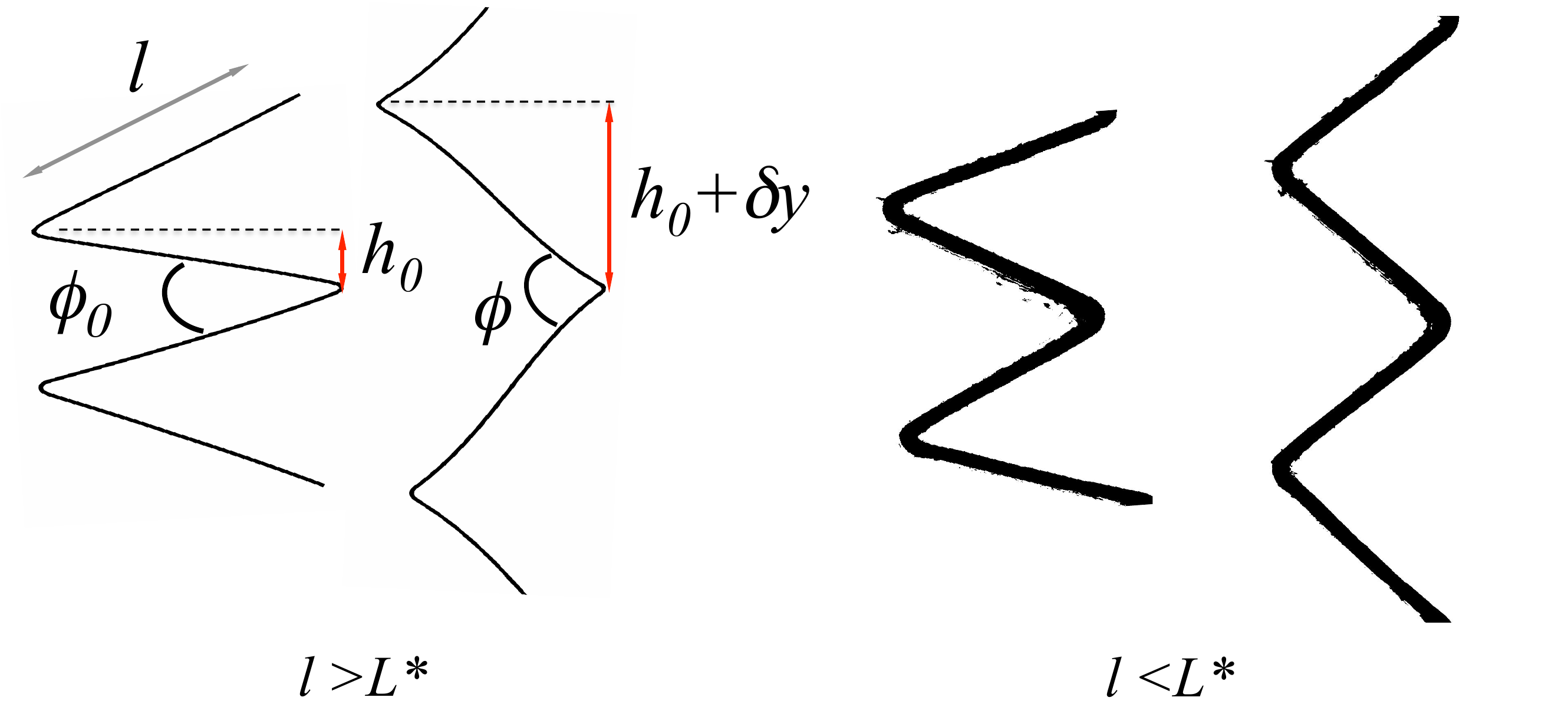}
\caption{Rest and deformed accordion-like origami made of folded Mylar sheets with $h=130 \mu m$ and for two different sizes of the panels. Left. $l= 2.5\mbox{cm} >L^*$, right  $l=0.6\mbox{cm}<L^*$.}
\label{fig:acc}
\end{figure}

The experimental results show that the crease stiffness is characterized by a torsional rigidity parameter $\kappa$ whose dimension is $N$, while the dimension of bending rigidity of the panels $B$ is $N.m$. Therefore, the ratio of these two parameters defines a length scale $L^*\equiv B/\kappa$, which in our case is a few centimeters (see Fig.~\ref{fig:Fig5}). In order to understand how these two modes of deformations compete, one can compute the variation of energy associated with the deformation of an origami structure around an equilibrium configuration.  The simplest origami is the ``accordion-like" folded strips, because it is controlled by two degrees of freedom, the rest angle $\phi_0$ of the crease and the length $l$ of panel (see Fig.~\ref{fig:acc}). Under a small applied tension, each element of the structure elongates by a total amount $\delta y$. The response of the hinge results in a contribution $\delta y_c$ to the total elongation given by $\delta y_c \approx (l/2)\cos\left(\phi_0/2\right)\delta\phi$, where $\phi = \phi_0+\delta\phi$ is the opening angle of the crease. As a consequence of the constitutive behavior of the crease, the associated energy cost reads $\delta E_c = \kappa W (\delta \phi)^2 =4\kappa W (\delta y_c)^2/\left(l^2\cos^2(\phi_0/2)\right)$. On the other hand, the bending deformation of the panel satisfies the linearized version of Eq.~\ref{eq:elastica}, which yields $\theta\left(s\right)=\theta_m\left(1-\left(2s/l\right)^2\right)$, where $\theta_m$ is an integration constant related to $\delta y_b $, the contribution to the total elongation $\delta y$. Using geometric arguments one finds that $\delta y_b/l \approx \frac{2}{3} \theta_m \cos\left(\phi_0/2\right)$. The bending energy of the panel can also be computed; it reads $E_b\approx 16 BW \theta_m^2/(3l)\approx 12 BW(\delta y_b)^2/\left(l^3\cos^2\left(\phi_0/2\right)\right)$. Finally, for a given fixed elongation $\delta y=\delta y_b + \delta y_c$ and to first order in $\delta\phi$, the minimization of $E_{tot}=E_b+E_c$ yields
\begin{equation}
\frac{\delta y_c}{\delta y}=\frac{1}{1+\frac{\kappa l}{3B}}=\frac{1}{1+\frac{l}{3L^*}}
\label{eq:lstar}
\end{equation}
Eq.~(\ref{eq:lstar}) states that for a given strip, when the linear panel size is small compared to the characteristic length scale $L^*$, the longitudinal extension of the structure is mostly due to the opening of the fold. In this limit, the panels can be considered as rigid. Thus, the unfolding mechanism corresponds to what would be expected for usual origami whose deformation along the creases can be deduced from kinematics~\cite{Wei2013}. In the opposite limit, when the linear size of the elementary panels is large compared to $L^*$, the extension of the structure is governed by the bending of these panels while the creases do not open up a lot. Fig.~\ref{fig:acc} shows the two opposite responses of a simple accordion like origami to an applied loading when the size of the panels is modified. Using simple arguments~\cite{deboeuf2013}, it is shown that the energy stored in the crease, and consequently the mechanical parameter $\kappa$, scales as $B/h$, which implies that the characteristic length scale $L^*$ is a linear function of the thickness. Fig.~\ref{fig:Fig5}  shows that the data are compatible with such a scaling, $L^* \simeq 200 h$. This large scale separation between $h$ and $L^*$ can be rationalized by noting that the bending modulus $B$ involves the Young modulus $E$, while the crease rigidity $\kappa$ involves plastic deformations, and therefore the yield stress $\sigma_Y$. The large dimensionless ratio of these two material parameters seems to govern this scale separation and thus the origami length scale $L^*$.

In summary, a mechanical characterization of creased sheets is presented, relying on a protocol that yields well defined reversible mechanical behavior and rest fold angle. Our technique provides an accurate measurement of the crease rigidity and the sheet's bending modulus $B$. We have shown that the rest angle and the origami length scale $L^*$ are governed by the ratio $\sigma_Y/E$ of the material, which appears as the relevant magnifying factor of the thickness in the origami response. 
This characteristic length scale $L^*$, which we expect to be a generic feature of more elaborate origami-like structures, represents a very simple design tool to predict the overall behavior of folded systems, and in particular wether they fold or bend. Since in general, it is of primary interest that the structure actually unfolds/actuates instead of bending/failing, this characterization also brings insights into the typical constraints on hinge mechanisms that should be associated with a given set of flexible panels to reach the desired function. 

\end{document}